\documentclass[aip,numerical,reprint,groupedaddress,showpacs,superscriptaddress,floatfix]{revtex4-1}  
\usepackage{graphicx,amsmath,amssymb,subfigure}
\newcommand{\bu}{{\bf u}}
\newcommand{\be}{{\bf e}}

\newcommand{\bufluc}{\tilde{\bu}}

\newcommand{\grad}{\nabla}
\newcommand{\lap}{\nabla^2}

\begin{document}
\title{Turbulent-laminar patterns in plane Poiseuille flow}
\author{Laurette S. Tuckerman}
\email{laurette@pmmh.espci.fr}
\affiliation{PMMH (UMR 7636 CNRS - ESPCI - UPMC Paris 6 - UPD Paris 7),
10 rue Vauquelin, 75005 Paris France}
\author{Tobias Kreilos}
\email{tobias.kreilos@epfl.ch}
\affiliation{Fachbereich Physik, Philipps-Universit\"at Marburg, 
35032 Marburg, Germany}
\affiliation{Max Planck Institute for Dynamics and Self-Organization,
Am Fassberg 17, 37077 G\"ottingen, Germany}
\affiliation{Emergent Complexity in Physical Systems Laboratory (ECPS), 
Ecole Polytechnique F\'ed\'erale de Lausanne, Switzerland}
\author{Hecke Schrobsdorff}
\email{hecke@nld.ds.mpg.de}
\affiliation{Max Planck Institute for Dynamics and Self-Organization,
Am Fassberg 17, 37077 G\"ottingen, Germany}
\author{Tobias M. Schneider}
\email{tobias.schneider@epfl.ch}
\affiliation{Max Planck Institute for Dynamics and Self-Organization,
Am Fassberg 17, 37077 G\"ottingen, Germany}
\affiliation{Emergent Complexity in Physical Systems Laboratory (ECPS), 
Ecole Polytechnique F\'ed\'erale de Lausanne, Switzerland}
\author{John F. Gibson}
\email{john.gibson@unh.edu}
\affiliation{Department of Mathematics and Statistics, 
University of New Hampshire, Durham NH 03824, USA}
\date{\today} 
 
\begin{abstract} 
Turbulent-laminar banded patterns in plane Poiseuille flow are studied 
via direct numerical simulations in a tilted and translating
computational domain using a parallel version of the 
pseudospectral code Channelflow. 
3D visualizations via the streamwise vorticity of an instantaneous and a 
time-averaged pattern are presented, as well as 2D visualizations of 
the average velocity field and the turbulent kinetic energy.
Simulations for $2300\geq Re_b \geq 700$ show the gradual
development from uniform turbulence to a pattern with wavelength 20 
half-gaps at $Re_b\approx 1900$, to a pattern with wavelength 40 at 
$Re_b\approx 1300$ and finally to laminar flow at $Re_b\approx 800$. 
These transitions are tracked quantitatively via
diagnostics using the amplitude and phase of the Fourier transform and 
its probability distribution. 
The propagation velocity of the pattern is approximately that of the 
mean flux and is a decreasing function of Reynolds number. 
Examination of the time-averaged flow shows that a turbulent band is 
associated with two counter-rotating cells 
stacked in the cross-channel direction and that 
the turbulence is highly concentrated near the walls.
Near the wall, the Reynolds stress force 
accelerates the fluid through a turbulent band while viscosity decelerates it; 
advection by the laminar profile acts in both directions.
In the center, the Reynolds stress force 
decelerates the fluid through a turbulent band 
while advection by the laminar profile accelerates it.
These characteristics are compared with those of turbulent-laminar 
banded patterns in plane Couette flow.

\end{abstract} 


\maketitle 

\section{Introduction: Phenomenon and Methods}
The transition to turbulence is one of the least understood 
phenomena in fluid dynamics. 
Transitional regimes in wall-bounded shear flows display 
regular patterns of turbulent and laminar bands which 
are wide and oblique with respect to the streamwise direction. 
These patterns have been studied in counter-rotating Taylor-Couette flow
\cite{Coles,Andereck,Hegseth89,Marques_PRE_09,Dong_PRE_09,Prigent_PRL,Prigent_PhysD}
and in plane Couette flow \cite{Prigent_PRL,Prigent_PhysD,Barkley_PRL_05,Barkley_JFM_07,Tuckerman_PF_11,Manneville_PRE_11,Duguet_JFM_10,Duguet_PRL_13,Brethouwer}.
\begin{figure}
\includegraphics[width=8cm]{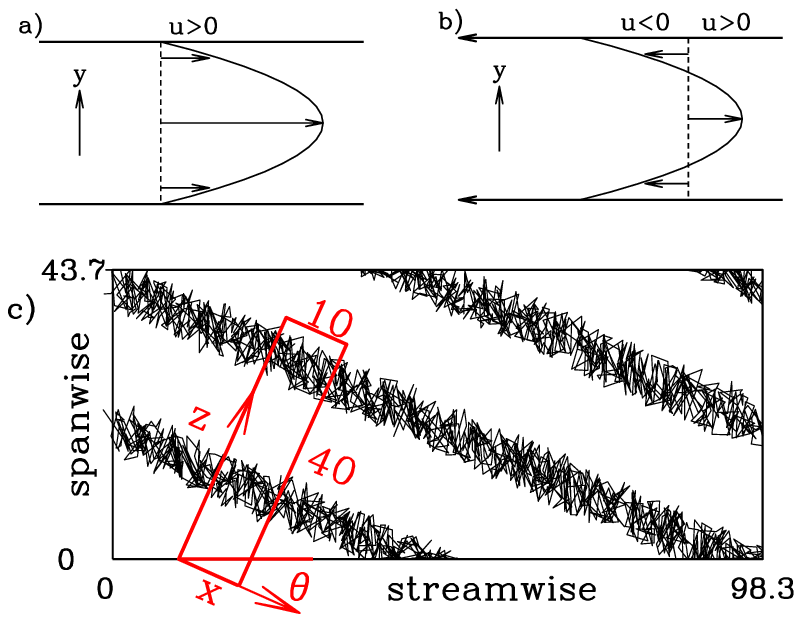}
\caption{
a) Standard (streamwise,cross-channel) view of plane Poiseuille flow 
in which $u_{\rm wall}\equiv u(y=\pm 1)=0$ and $u_{\rm bulk}\equiv \frac{1}{2}\int_{-1}^{1} u(y) dy = \frac{2}{3}$. b) Translating reference frame such that $u_{\rm wall}=-\frac{2}{3}$ and $u_{\rm bulk} = 0$. c) Tilted reference frame in 
(streamwise, spanwise) view where $x$ is aligned with 
schematically drawn turbulent bands, at an angle of $\theta=24^\circ$ 
to the streamwise direction, and $z$ aligned with the pattern wave-vector.
The tilted box has dimensions $L_x\times L_z=10\times 40$.
In order to capture the same pattern, a box aligned with the streamwise 
and spanwise directions would be required to have dimensions 
$L_{\rm strm}\times L_{\rm span} = 40/\sin(24^\circ) \times
40/\cos(24^\circ)=98.3\times 43.7$.}
\label{fig:domain}
\end{figure}
\begin{figure*}
\includegraphics[width=12cm,clip]{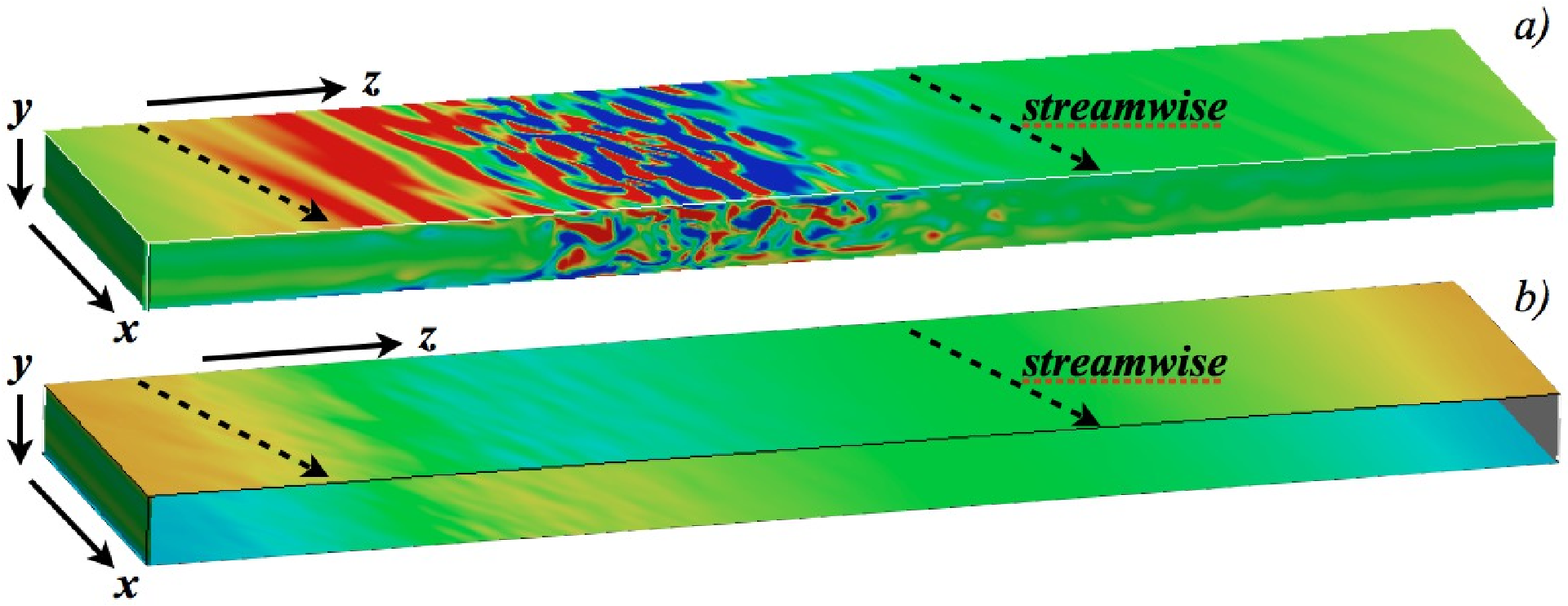}
\caption{3D visualization of the a) instantaneous and b) time-averaged
$(\Delta T=8000)$ streamwise vorticity
of a turbulent-laminar banded state at $Re=1100$ over the entire domain of size 
$L_x\times L_y\times L_z= 10\times 2 \times 40$.
The bands are parallel to the $x$ direction, while the 
small-scale vortices are aligned in the streamwise direction (dashed arrows) 
which is oriented at an angle of 24$^\circ$ to the $x$ direction.
The maximum and minimum of the instantaneous
streamwise vorticity is $\pm 6.2$ while that of the time-averaged 
vorticity is only $\pm 0.32$. Features of both are emphasized by choosing 
the color scale $[-0.5,0.5]$. The instantaneous vorticity, 
while strongest on the bounding plates, is present throughout 
the domain, as shown by the $(y,z)$ plane in front.
This plane is omitted from the time-averaged vorticity to
reveal contrasting views on the upper and lower plates; 
the antisymmetry in the vorticity results from 
the $y$-reflection symmetry of the average velocity.}
\label{fig:3D_view}
\end{figure*}

Turbulent-laminar banded patterns have also been observed 
numerically and experimentally in 
plane Poiseuille (channel) flow by Tsukahara et al.~\cite{Tsukahara_TSFP_05,Tsukahara_THMT_06,Tsukahara_ASCHT_07,Hashimoto_THMT_09}. 
Tsukahara et al.~\cite{Tsukahara_THMT_06} presented detailed visualizations 
from numerical simualtions of the mean flow as well as the effect of 
turbulent bands on heat transport. 
Later experiments~\cite{Hashimoto_THMT_09} compared the range of 
Reynolds numbers, wavelengths and angles of the turbulent bands obtained 
experimentally with the numerical results reported by 
Tsukahara et al.~\cite{Tsukahara_ASCHT_07}.
Brethouwer et al.~\cite{Brethouwer} simulated a turbulent-laminar 
pattern in Poiseuille flow as part of a larger study 
investigating the effects of damping by Coriolis, buoyancy and Lorentz 
forces on patterns in transitional flows. 
The goal of the present paper is to extend this work using 
methods previously employed to study turbulent-laminar banded patterns 
in plane Couette flow. 
In particular, we wish to determine if these patterns can be reproduced 
in the minimal geometry used in simulations of plane Couette flow 
\cite{Barkley_PRL_05,Barkley_JFM_07,Tuckerman_PF_11}, 
and to describe their evolution in time, their propagation velocity, 
and the balance of forces they entail.


Plane Poiseuille flow is generated by an imposed pressure gradient 
or an imposed bulk velocity between two parallel rigid plates. The length scale
for nondimensionalization is half the distance between the plates. 
For a velocity scale, several choices are common, leading to several 
definitions of the Reynolds number: $Re_c$ uses the velocity at the 
center of the channel, $Re_b$ uses the bulk velocity,
and $Re_\tau$ uses the wall shear velocity. One standard choice, and that made 
here, is to impose the bulk velocity and to scale 
the velocity by $3 u_{\rm bulk}/2$, because this leads to $Re_b=Re_c$ for 
the laminar flow. Many of the references we cite use 
another factor, one~\cite{Brethouwer} or 
two~\cite{Tsukahara_TSFP_05,Tsukahara_THMT_06,Tsukahara_ASCHT_07,Hashimoto_THMT_09,Kim}, 
in place of the 3/2; when we cite Reynolds numbers from these references
we have multiplied them by the appropriate conversion factor of 3/2 or 3/4.
Unless mentioned otherwise, the Reynolds number $Re_b$ is denoted merely by 
$Re$. 
The computational domain used in this study is tilted with respect to 
the bulk velocity, as illustrated in Fig.~\ref{fig:domain}c, 
in order to efficiently capture similarly tilted laminar-turbulent
patterns.
As was done for plane Couette flow
\cite{Barkley_PRL_05,Barkley_JFM_07,Tuckerman_PF_11}, 
the horizontal part of the domain is a narrow rectangle whose short direction 
(here, the $x$ axis, with $L_x=10$) is parallel to the expected direction of 
the bands, at an angle of 24$^\circ$ from the streamwise direction. 
The long direction (here, the $z$ axis, with $L_z=40$) is parallel to the 
expected wave-vector of the bands. Thus:
\begin{subequations}
\begin{eqnarray}
\mathbf{\hat{e}}_{\rm strm} &=& 
\cos 24^\circ\:\mathbf{\hat{e}}_x+\sin 24^\circ\:\mathbf{\hat{e}}_z,\\
\mathbf{\hat{e}}_{\rm span} &=
-&\sin 24^\circ\:\mathbf{\hat{e}}_x+\cos 24^\circ\:\mathbf{\hat{e}}_z
\end{eqnarray}\label{eq:tilted}\end{subequations}
The reason for choosing $24^\circ$ in plane Couette flow was that this angle 
is in the range $[24^\circ,37^\circ]$ observed experimentally in very 
large-scale experiments \cite{Prigent_PRL,Prigent_PhysD} 
(770 by 340 half-gaps in the streamwise and 
spanwise directions, respectively), in which the flow was free to 
choose its own angle; this range of angles is also observed in 
simulations by Duguet et al.~\cite{Duguet_JFM_10} in a domain of similar size. 
This is also the case for plane Poiseuille flow. 
Tsukahara et al.~\cite{Tsukahara_TSFP_05,Tsukahara_THMT_06}
first produced turbulent-laminar patterns in a domain 
with dimensions of 51.2 by 22.5 in the streamwise and 
spanwise directions, leading by construction 
to $\theta=\tan^{-1}(22.5/51.2)=23.7^\circ$.
The domain used by Brethouwer et al.~\cite{Brethouwer} is very similar 
(55 by 25) and hence leads to a similar angle of $24.4^\circ$.
Later simulations~\cite{Tsukahara_ASCHT_07} in a domain of size 
328 by 128 (in which the flow was relatively free to 
choose its own angle) produced patterns with angles in the 
range $[20^\circ,25^\circ]$ while the experiments of 
Tsukahara et al.~\cite{Hashimoto_THMT_09} 
showed angles in the range $[20^\circ,30^\circ]$.
Our narrow tilted domain enforces an angle of $24^\circ$; 
only patterns with this angle can be simulated. 

The tilted $x$ and $z$ directions are taken to be periodic and 
$y$ is the usual cross-channel direction. 
The streamwise, cross-channel and spanwise velocities continue to be 
denoted by ${\bf u}=(u,v,w)$ (even though $x,y,z$ do not correspond to these
directions). 
In order to follow the patterns as they advect with the bulk flow,
computations are performed in a moving reference frame whose velocity
matches the constant (nondimensionalized) streamwise bulk velocity of 
2/3. In this reference frame the walls move at $-2/3$ in the 
streamwise direction, and the imposed mean velocity is zero in both the 
span and streamwise directions. All velocities in this study are reported 
with respect to the moving reference frame (except those involved in the 
definition of Reynolds numbers, which are relative to fixed walls). 
The domain size is
$L_x\times L_y\times L_z=10 \times 2 \times 40$. 
The choice of $L_z$ was guided by considerations similar to 
those for the angle, i.e. results from experiments and simulations in 
plane Couette flow and plane Poiseuille flow. 
All of the references cited previously
~\cite{Brethouwer,Tsukahara_TSFP_05,Tsukahara_THMT_06,Tsukahara_ASCHT_07,Hashimoto_THMT_09} reported patterns with 
wavelengths in the range $[20,30]$. In our domain, only 
patterns whose wavelength is a divisor of $L_z$ can be simulated, 
i.e. 40, 20, 10, etc. 
The choice $L_x=10$ is dictated by the requirement 
that the box be large enough to sustain turbulence, 
more specifically that the spanwise dimension $L_x\sin\theta$ 
be wide enough to accomodate a pair of streamwise vortices~\cite{Jimenez_91,Hamilton,Waleffe_03,Barkley_PRL_05,Barkley_JFM_07,Tuckerman_PF_11}. 

Streamwise vortices are indeed a prominent feature of turbulent regions, 
as shown in the visualization in Fig.~\ref{fig:3D_view}a of a 
computed turbulent-laminar pattern. 
The streamwise vorticity is particularly appropriate for 
representing turbulence in plane Poiseuille flow since it is zero 
for laminar flow and is not zero at the plates, 
near which the turbulence is most intense. 
The instantaneous vorticity, Fig.~\ref{fig:3D_view}a, 
is localized in one region of the domain 
and is aligned in the streamwise direction. 
Figure \ref{fig:3D_view}b shows the streamwise vorticity averaged over 
$\Delta T=8000$ time units. 
The time-averaged vorticity is much weaker than the instantaneous vorticity, 
and is antisymmetric under reflection in $y$, showing that the 
corresponding velocity is reflection-symmetric in $y$.

\begin{figure*}
\includegraphics[width=16cm]{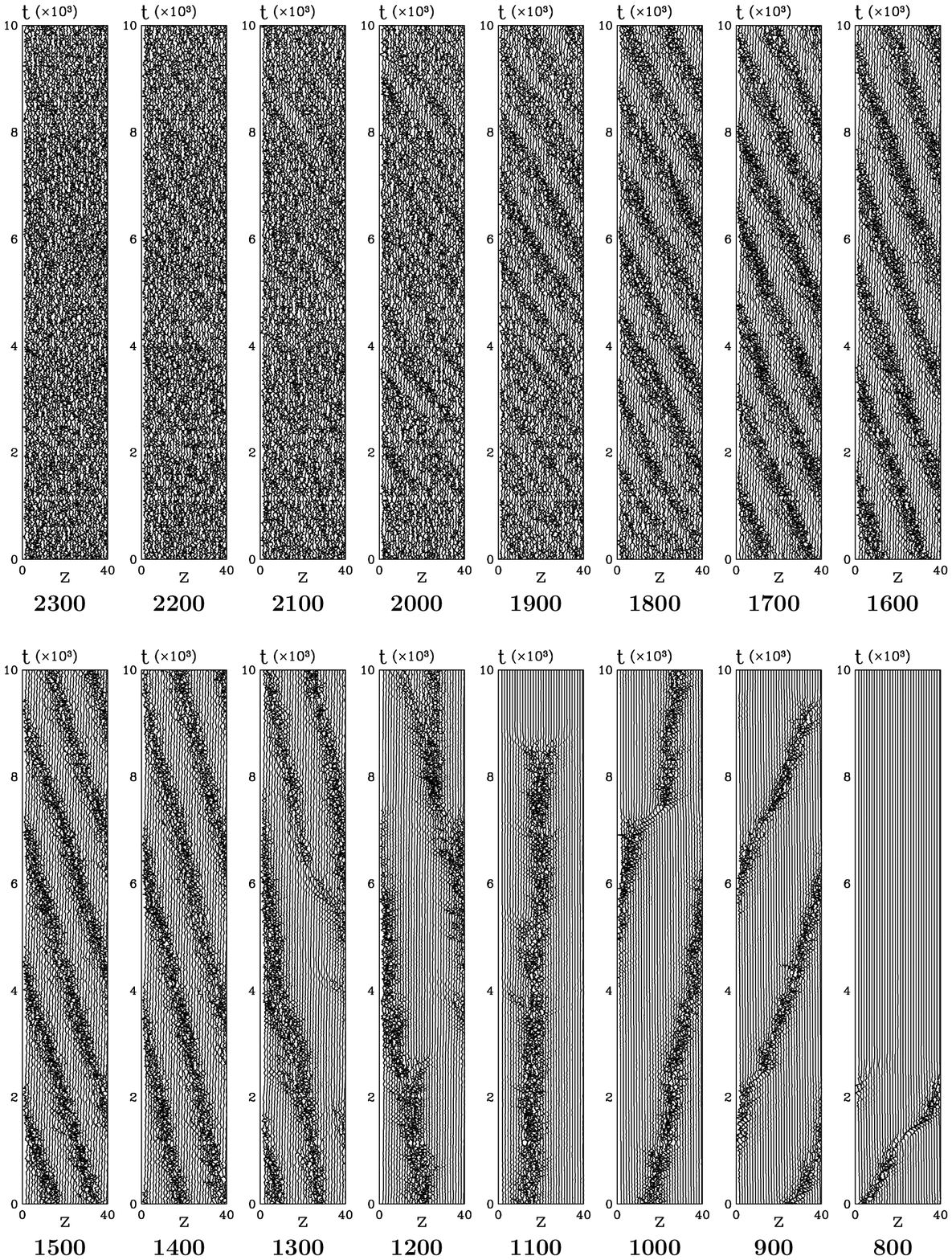}
\caption{Spatio-temporal plots from simulations in domain of size 
$L_x\times L_y \times L_z = 10 \times 2 \times 40$ 
in which the $z$ direction is oriented along the pattern wave-vector.
Spanwise velocity timeseries $w(z_j,t)$ are shown for points
along the line $x=0$, $y=0.8$ at 32 equally spaced values $z_j=j L_z/32$
for $2300 \geq Re \geq 800$. Uniform turbulence can be seen 
for $Re=2300$, traces of laminar patches 
for $2200 \geq Re\geq 2000$, increasingly well-defined left-going bands 
of wavelength 20 for $1900 \geq Re \geq 1400$, 
a change in wavelength and direction for $Re=1300$, 1200, 
and an increasingly weak and fragile right-going pattern of 
wavelength 40 for $1100 \geq Re \geq 800$.}
\label{fig:longtime}
\end{figure*}

The simulations were performed with a parallelized version of the 
pseudospectral C++-code Channelflow \cite{channelflow}, which employs
Fourier-Chebyshev spatial discretization, fourth-order semi-implicit 
backwards-differentiation time stepping, and an influence matrix method
with Chebyshev tau correction on the primitive-variables formulation
of the Navier-Stokes equations~\citep{Orszag,Kleiser,Kim,Peyret}.
The code uses FFTW~\cite{Frigo2005} for Fourier transforms and MPI 
for parallelization. 
We use 
$N_x\:\times\:(N_y+1) \:\times\:N_z=128\;\times\;65\;\times\;512 =4.2\times 10^6$ 
points or modes to represent the 
domain of size $L_x\times L_y\times L_z=10 \times 2 \times 40$, 
with a spacing of $\Delta x=\Delta z = 0.08$ and $\Delta y$ ranging from 
$\Delta y_{\rm wall}=1-\cos(\pi/64)=0.001$ 
to $\Delta y_{\rm mid}=\cos(31\pi/64)=0.05$. 
For the highest Reynolds number we simulate, $Re=2300$, 
the ratio between the viscous wall unit and the half-gap is
0.009, so $\Delta x^+ = \Delta z^+=0.08/0.009=9$, 
$\Delta y_{\rm wall}^+=0.001/0.009=0.11$ and 
$\Delta y_{\rm mid}^+=0.05/0.009=5.6$.
For plane Poiseuille flow, it is crucial to have sufficient resolution 
in the cross-channel direction near the walls, where the turbulence 
is concentrated. 
The resolution used here is similar to that used in the simulations 
by Kim et al.~\cite{Kim} ($Re= 4200$; $Re_\tau=180$) 
and Jim\'enez \& Moin \cite{Jimenez_91}; 
although these authors used $N_y=128$, they studied Reynolds
that were about twice the highest Reynolds number investigated here and 
stated that using $N_y=64$ produced similar results \cite{Kim}. 
In terms of wall units, our resolution is the same or finer than that 
used in these studies. Our resolution is also close to that used by 
Tsukahara et al. \cite{Tsukahara_TSFP_05,Tsukahara_THMT_06} for
$Re=1732$ ($Re_\tau=80$) and $Re=1327$ ($Re_\tau = 64$) 
and higher than that of  
Brethouwer et al.~\cite{Brethouwer}, who used 
$\Delta x^+=15$, $\Delta z^+=6.7$ and $N_y=32$ for a case 
with $Re=700$ ($Re_\tau=69$). 
The timestep varied from $\Delta t=0.03$ for $Re=900$ to 
$\Delta t = 0.015$ for $Re=2300$.
Simulations run on the IBM x3750 of the IDRIS supercomputer
center using 32 processes typically took about 16 wall clock 
hours to simulate 10 000 advective time units. 

 
\section{Reynolds-number scan}
Figure \ref{fig:longtime} shows spatio-temporal diagrams 
of the spanwise velocity. 
For $Re\leq 2000$, each simulation is a continuation of 
the corresponding part of a long simulation in which the 
Reynolds number is decreased in discrete decrements of 100, 
which will be presented in Fig.~\ref{fig:time} 
and which itself is initialized with random noise. 
The simulations with $Re > 2000$ were all initialized with the 
final state of the $Re=2000$ run.
The timeseries show $w(z_j,t)$ at $x=0$, $y=0.8$ (near the upper plate), 
for 32 values $z_j$ separated by intervals of $\Delta z = L_z/32$ 
for Reynolds numbers varying from 2300 down to 800. 
Dark patches indicate rapid large-amplitude 
oscillations in the spanwise velocity, i.e. turbulent regions. 
The surrounding lighter patches are composed of straight lines, 
indicating locations at which the spanwise velocity remains constant 
or nearly so, i.e. quasi-laminar regions. 

For the highest Reynolds number, $Re=2300$, 
the entire interval $0 \leq z \leq L_z$ is dark, 
indicating turbulence which is statistically uniform over the domain. 
As $Re$ is lowered to 2100, quiescent patches appear 
which move towards the left (opposite to $(u_{\rm mean}-u_{\rm wall})$).
Timeseries for $1900 \geq Re \geq 1400$ show two clearly delineated 
turbulent bands with a fairly well-defined wavelength and velocity. 
From $Re=1300$ to $Re=1200$, there is a transition from two to one turbulent 
band. New turbulent patches repeatedly branch off from existing ones; 
these are more persistent and long-lasting for $Re=1300$ than for 1200. 
The velocity of the pattern decreases. 
For $Re=1100$, the pattern comprises a single band and is almost stationary;
the band begins to disappear at $T=8700$, becoming 
completely laminar by $T=9000$.
For $Re=1000$ and 900, a single right-going band is present, 
which, for $Re=900$ disappears at $T=9700$. For $Re=800$, 
the band disappears earlier, at around $T=2200$. 

\begin{figure*}
a)\includegraphics[width=2.5cm]{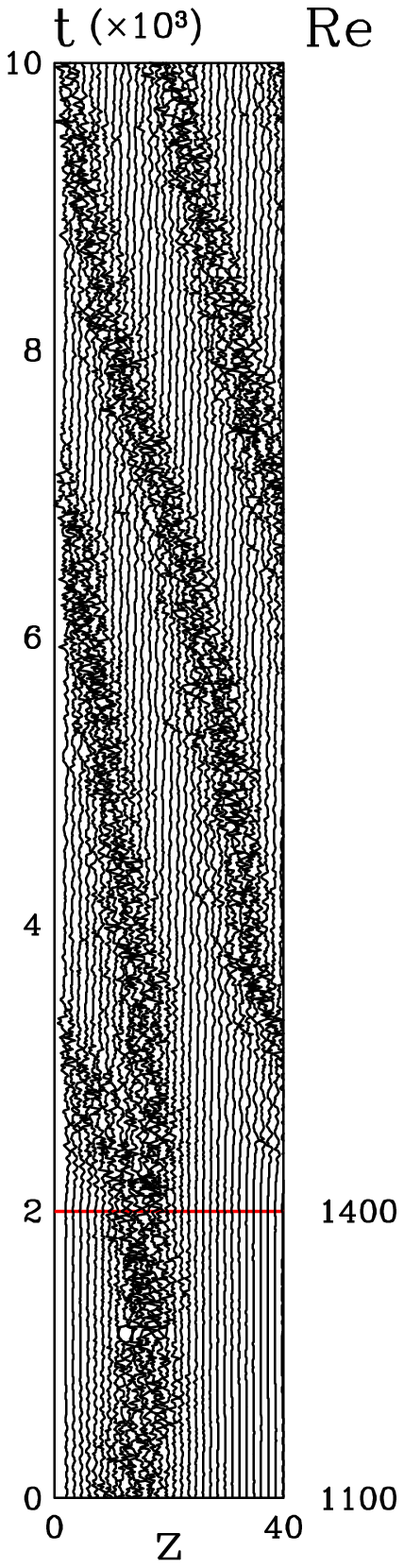}
b)\includegraphics[width=9cm]{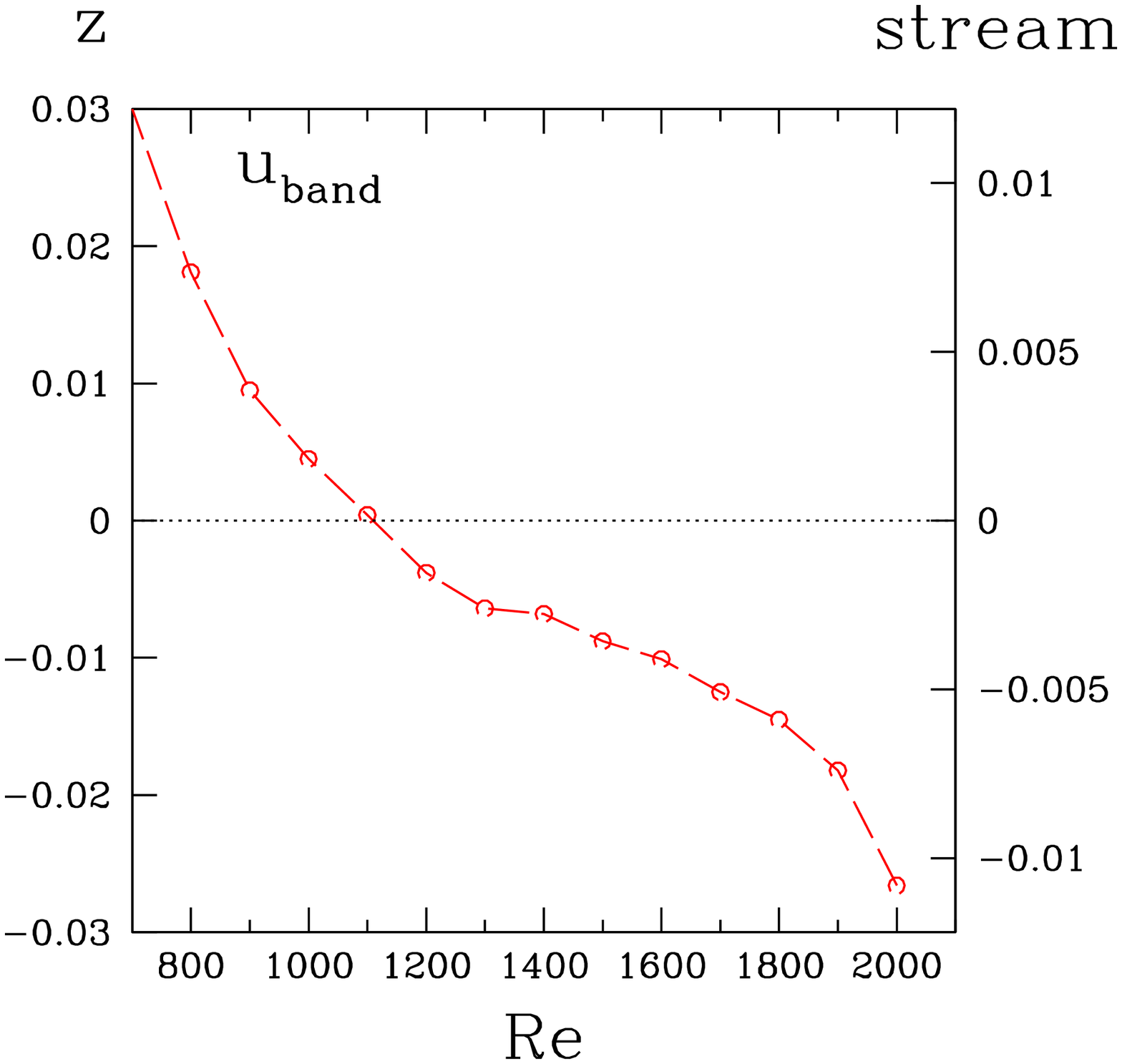}
\caption{
a) Simulation in domain of size 
$L_x\times L_y \times L_z = 10 \times 2 \times 40$, 
with conventions as in Fig.~\ref{fig:longtime}.
When $Re$ is increased from 1100 to 1400, 
the pattern with one turbulent band is quickly replaced 
by a left-moving pattern with two bands.
Propagation speed of turbulent bands with respect to the mean flow.
At $Re\approx 1100$, the bands move at approximately the same speed 
as the mean flow. For $Re\lesssim 1100$, the bands move more quickly than the 
mean flow, while for $Re \gtrsim 1100$, they move more slowly. 
The scale on the left (right) indicates the velocity in the 
$z$ (streamwise) direction; the two scales are related by the trigonometric
factor $\sin 24^\circ$.}
\label{fig:velocities}
\end{figure*}

For plane Couette flow, a qualitative distinction can be made 
between patterns at higher $Re$, in which the turbulent and quasi-laminar 
regions each occupy approximately half of the domain, 
and patterns at lower $Re$, in which the turbulent band occupies 
a smaller fraction of the domain. 
The averaged high-$Re$ patterns were shown~\cite{Barkley_JFM_07}
to have a trigonometric dependence on $z$.
In contrast, the bands at lower $Re$ were shown to be isolated states, 
in that they retain their size when placed in a wider 
domain~\cite{Barkley_PRL_05} and are surrounded by truly laminar regions. 
A comparison of the states at $Re\geq 1400$ 
with those at $Re \leq 1100$ shows that this distinction seems also to 
apply to plane Poiseuille flow. 
For $Re\geq 1400$, the turbulent bands occupy about half the width of 
the domain, with the other half consisting of slightly chaotic flow, 
as shown by the small-scale oscillations in $w(z_j,t)$. 
For $Re\leq 1100$, the single turbulent band occupies much less than the width of 
the domain and the flow reverts to laminar quite close to the 
boundaries of the band, as shown by the straight lines $w(z_j,t)$ 
for $z_j$ within a few multiples of $L_z/32$.

The branching events at $Re=1300$ and $Re=1200$ indicate bistability 
between a pattern with wavelength 40 and wavelength 20;  
a domain with a larger or different $L_z$ would almost surely display 
patterns with intermediate wavelengths and 
bistability at a different value of $Re$.
There is also clearly an important random component in 
the fact that relaminarisation occurs at $Re=1100$ but not at $Re=1000$. 
Simulations with different initial conditions would almost surely 
lead to relaminarisation at different times; 
the properties of these events must be studied statistically, 
as has been done for pipe flow~\cite{Peixinho_JFM_07,Avila_Science} 
and for Couette flow~\cite{Shi}.
Although we have not checked systematically for hysteresis, 
when we increased the Reynolds number from 1100 to 1400, 
the initial single quasi-stationary band 
evolved quickly to a left-moving pattern with two bands; 
see figure 
\ref{fig:velocities}a).

Figure \ref{fig:velocities}b) shows the propagation velocity as a function of
$Re$.  Propagation in the $z$ direction is to be expected, since there is a
substantial overlap between the $z$ direction and the streamwise
direction. The fact that the speeds are so small demonstrates that the
turbulent bands move essentially at the speed of the mean flow, as was noted
by Tsukahara et al.~\cite{Tsukahara_THMT_06,Hashimoto_THMT_09}.  
As was seen in Fig.~\ref{fig:longtime}, 
$Re\approx 1100$ separates propagation to the left (slower than the mean flow) 
and to the right (faster than the mean flow).  It is because the pattern 
at $Re=1100$ is approximately stationary in the frame of the mean flow
that this Reynolds number was chosen to display the time-averaged flow shown 
in Fig.~\ref{fig:3D_view}.
Although the number of bands is quantized, their velocity is not; 
moreover the velocity varies smoothly through the change in the 
number of bands at $Re=1100$. Therefore it seems likely that 
that the velocity presented in Fig.~\ref{fig:velocities}b) is independent 
of the domain.
Propagation velocities are in general quite sensitive to resolution; 
previous simulations with less resolution in $y$ showed the propagation 
velocity to change sign at $Re=1400$ instead of $Re=1100$. 
We have verified that 
the velocity does not change appreciably with higher resolution. 

An overall view of the evolution of the pattern can be seen in 
the Reynolds-number scan of Fig.~\ref{fig:time}. 
This figure describes a simulation 
initialized at $Re=2000$ with a random initial condition 
and in which the Reynolds number is 
lowered in discrete steps of 100, remaining at 
each $Re$ for a time of length $\Delta T=2000$. 
(The simulations of length $\Delta T=10\,000$ in Fig.~\ref{fig:longtime} are 
continuations for another $\Delta T=8000$ of each of the sections of 
Figure \ref{fig:time}). 
Figure \ref{fig:time}a shows streamwise velocity profiles $u(y)$
at intervals of $\Delta T=2000$ at a fixed value of $x$ and $z$; 
these are flat or parabolic, depending on whether the 
corresponding flow or region is turbulent or laminar. 
Figure \ref{fig:time}b, like Fig.~\ref{fig:longtime}, 
shows timeseries of the spanwise velocity $w(z_j,t)$ at 
32 equally spaced points in $z$. 

The evolution from the random initial condition at $Re=2000$ leads rapidly 
to turbulence which is uniform (without bands).
As $Re$ is lowered past $Re\approx 1800$, two quiescent patches appear 
(though the long time series of Fig.~\ref{fig:longtime} show that a 
muted version of this pattern already appears for higher $Re$, 
given sufficient time). 
Several transitions are clearly visible: 
from two turbulent bands to one at $Re=1200$, 
from leftwards to rightwards motion at $Re=1100$, and from 
one turbulent band to none at $Re=700$.

Figures \ref{fig:time}c,d show that these tendencies can be measured 
quantitatively via the modulus $\vert\hat{w}_m(t)\vert$ and 
the phase $\hat{z}_m(t)$ of the discrete Fourier transform in $z$:
\begin{equation}
w(z_j,t)=\sum_m \hat{w}_m(t) e^{imz_j\frac{2\pi}{L_z}}
=\sum_m \vert\hat{w}_m(t)\vert e^{im(z-\hat{z}_m(t)) \frac{2\pi}{L_z}}
\vspace*{-0.2cm}
\label{eq:inst}\end{equation}
averaged over appropriate time intervals:
\begin{subequations}
\begin{eqnarray}
\langle\vert\hat{w}_m(t)\vert\rangle&_{1000}\equiv&\frac{1}{1000}
\int_{t^\prime=0}^{1000} dt^\prime\;\vert\hat{w}_m(t+t^\prime)\vert \\
\langle \hat{z}_m(t)\rangle_{100}&\equiv&\frac{1}{100}\int_{t^\prime=0}^{100} 
dt^\prime\;\hat{z}_m(t+t^\prime)
\end{eqnarray}
\label{eq:tavg}\end{subequations}
In Fig.~\ref{fig:time}c, $\langle\vert\hat{w}_2(t)\vert\rangle_{1000}$ 
rises from a low value when $Re$ is decreased below 1900, and is then overtaken 
by $\langle\vert\hat{w}_1(t)\vert\rangle_{1000}$ at $Re=1200$, 
when one of the turbulent bands disappears. 
Although the pattern for $1200 \geq Re \geq 800$ has wavelength $\lambda=40$, 
it also contains higher harmonics and so $\hat{w}_2$ remains non-negligible.

Figure \ref{fig:time}d shows the averaged phases 
$\langle \hat{z}_m(t)\rangle_{100}$.
When the modulus is small, the phase loses significance, 
so phases are shown only when the modulus exceeds a heuristically determined 
threshold, 0.24 for $m=1$ and 0.4 for $m=2$.
The phases $\langle \hat{z}_2(t)\rangle_{100}$ and 
$\langle \hat{z}_2(t)\rangle_{100}+L_z/2$ track 
the centers of the two turbulent bands seen in Fig.~\ref{fig:time}b 
for $1800 \geq Re > 1200$, while $\langle \hat{z}_1(t)\rangle_{100}$ tracks the 
center of the single turbulent band for $1200 \geq Re \geq 800$.

The instantaneous values of 
$a(t)\equiv\vert\hat{w}_m(t)\vert$ collected for each Reynolds number 
during the time that the flow is partly or entirely turbulent 
(see Fig.~\ref{fig:longtime}) 
can also be used to construct probability distribution functions. 
Because $\vert\hat{w}_m(t)\vert$ is a modulus, the range 
$a_{j-1} < \vert\hat{w}_m(t)\vert \leq a_j$ corresponds to an annulus in the two-dimensional 
Cartesian space of $(\hat{w}_m^r,\hat{w}_m^i)$ of area 
\begin{equation}
\pi (a_j^2-a_{j-1}^2)= 2\pi \frac{(a_j+a_{j-1})}{2} (a_j - a_{j-1})
\label{eq:norm} \end{equation}
The bin boundaries $a_j$ can be chosen to correspond to 
annuli of equal size by taking
\begin{subequations}
\begin{eqnarray}
a_j &\equiv& \sqrt{\frac{j}{N_{\rm bin}}} \max_t \vert\hat{w}_m(t)\vert\\ 
p_j &\propto& \vert\{t: a_{j-1} < \vert\hat{w}_m(t)\vert \leq a_j\}\vert
\end{eqnarray}
\end{subequations}
where $\vert\{\}\vert$ denotes the number of elements of a set.
Another possibility is to choose bin boundaries $a_j$ which are equally 
spaced and to correct for the difference in annular areas 
(\ref{eq:norm}) by dividing $p_j$ by $(a_j+a_{j-1})/2$. 
A third possibility is to choose bin boundaries such 
that each bin contains the same number of values, and to 
divide $p_j$ by $(a_j-a_{j-1})(a_j+a_{j-1})/2$. 
All three procedures lead to similar probability distribution functions.

\begin{figure*}
\includegraphics[width=14cm]{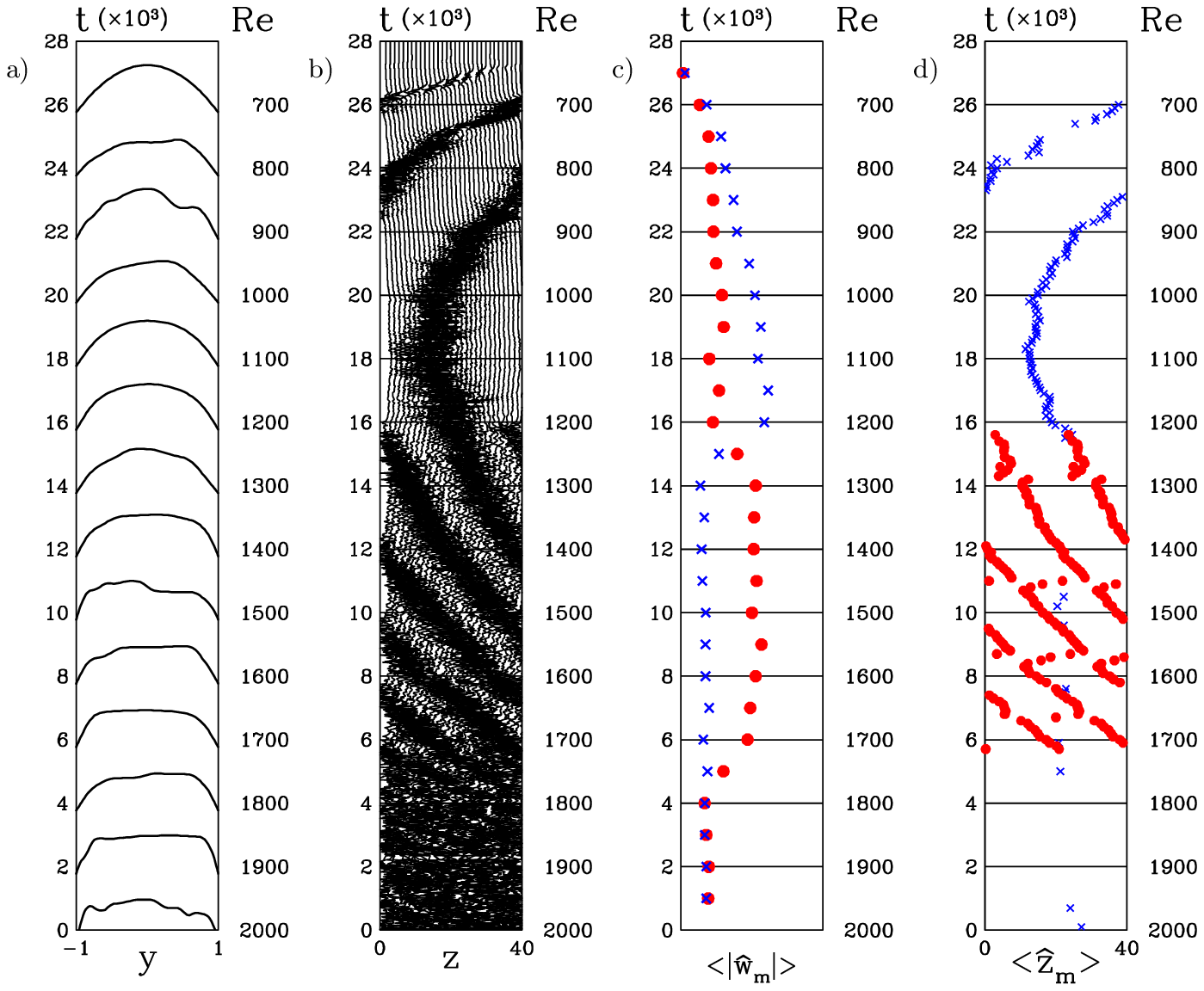}
\vspace*{-0.5cm}
\caption{Simulations in domain of size 
$L_x\times L_y \times L_z = 10 \times 2 \times 40$ 
The Reynolds number is decreased from 2000 to 700
in discrete steps at time intervals of $\Delta T=2000$, 
as indicated along the axes on the right. \newline
a) Instantaneous representative streamwise velocity profiles $u(y)$ along the line $x=z=0$ at intervals of $\Delta T=2000$.\newline
b) Spanwise velocity timeseries $w(z_j,t)$ along the line $x=0$, $y=0.8$ 
at 32 equally spaced values $z_j=j L_z/32$.
\newline
c) Temporal average $\langle\vert\hat{w}_m(t)\vert\rangle_{1000}$ 
(arbitrary units) 
of the modulus of the $z$-Fourier transform $\hat{w}_m$ of the 
spanwise velocity.
\newline
d) Temporal average $\langle \hat{z}_m(t)\rangle_{100}$ 
of the phase of $\hat{w}_m$ at times for which 
$\langle\vert\hat{w}_m(t)\vert\rangle_{100}$ is sufficiently large.
\newline
For c),d), the red disks indicate $m=2$
($\lambda=20$), while the blue crosses indicate $m=1$ ($\lambda=40$). 
}
\label{fig:time}
\vspace*{0.5cm}
\includegraphics[width=16cm]{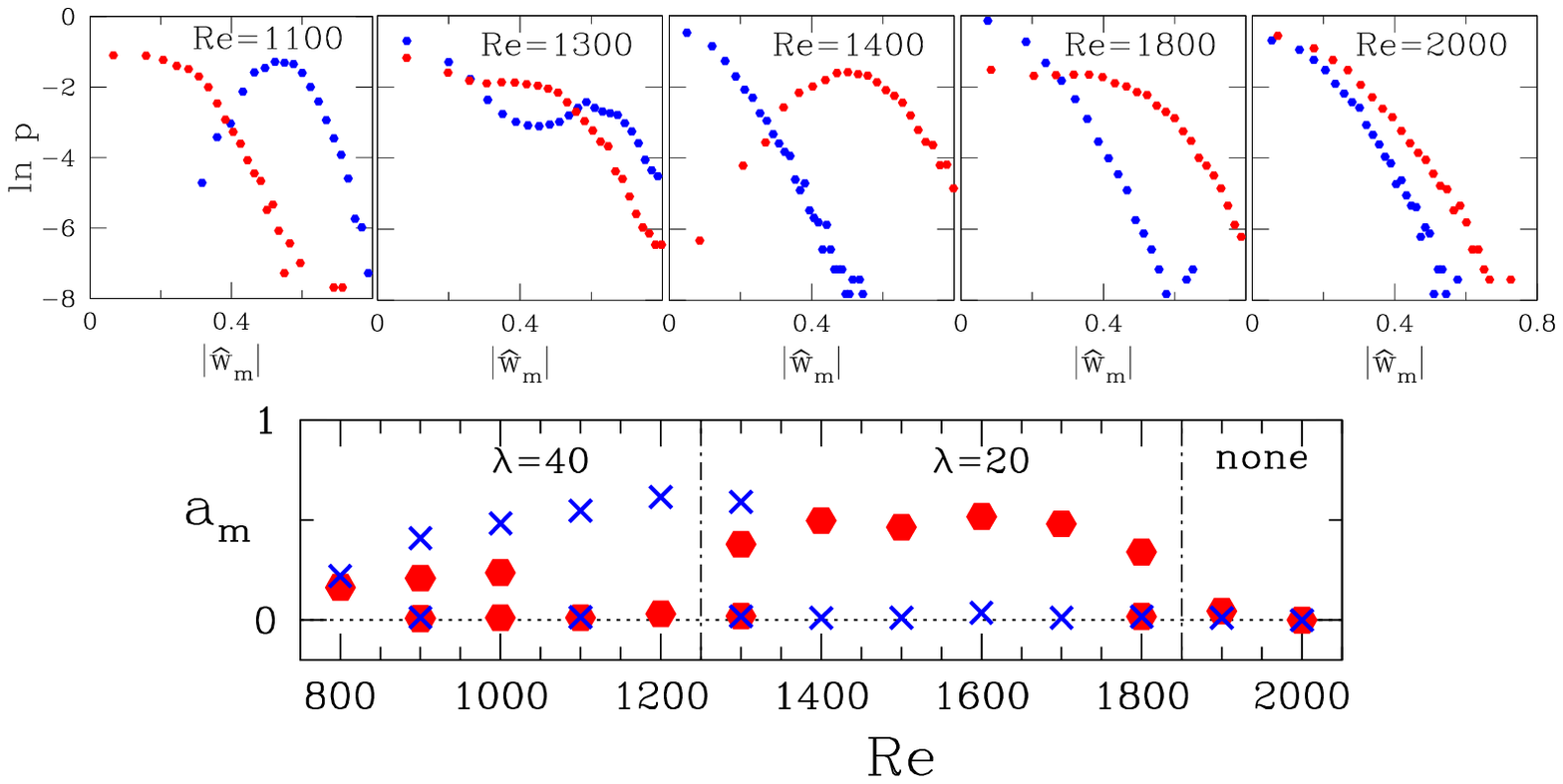}
\vspace*{-0.4cm}
\caption{Above: probability distributions of the moduli of the Fourier 
components $\vert\hat{w}_m(t)\vert$ for $m=1$ and $m=2$ for representative 
values of the Reynolds number.
Below: maxima of the PDFS as a function of Reynolds number 
showing changes in regime at $Re\approx 1900$ and $Re\approx 1300$.
The red disks indicate $m=2$ ($\lambda=20$), while the blue crosses 
indicate $m=1$ ($\lambda=40$).
The maximum of both PDFS is located at zero for $Re\geq 1900$.
For $1900 > Re \geq 1300$, the $m=2$ PDF 
has a maximum away from zero, indicating a pattern of this wavenumber. 
For $1300 \geq Re \geq 800$, the $m=1$ PDF 
has a maximum away from zero, indicating a pattern of this wavenumber.}
\label{fig:pdfs_arrange}
\end{figure*}

Figure \ref{fig:pdfs_arrange} displays probability distribution functions
for $m=1$ and $m=2$ for representative Reynolds number values.
In the absence of a pattern, in particular for uniform turbulence, 
the maximum (most probable value) for $\vert\hat{w}_m(t)\vert$ is zero, 
while for patterned flows, the maximum is non-zero. 
The PDFs yield thresholds:\\
$\bullet$ $Re\approx 1900$, separating uniform turbulence and a pattern with $m=2$, 
i.e. $\lambda=20$\\
$\bullet$ $Re\approx 1300$, separating patterns with $m=2$ and $m=1$, 
i.e. $\lambda=20$ and $\lambda=40$\\
$\bullet$ $Re \approx 800$, separating a pattern with $m=1$ from laminar Poiseuille flow. \\

\section{Mean flow and force balance}

\begin{figure}
\includegraphics[width=8.8cm]{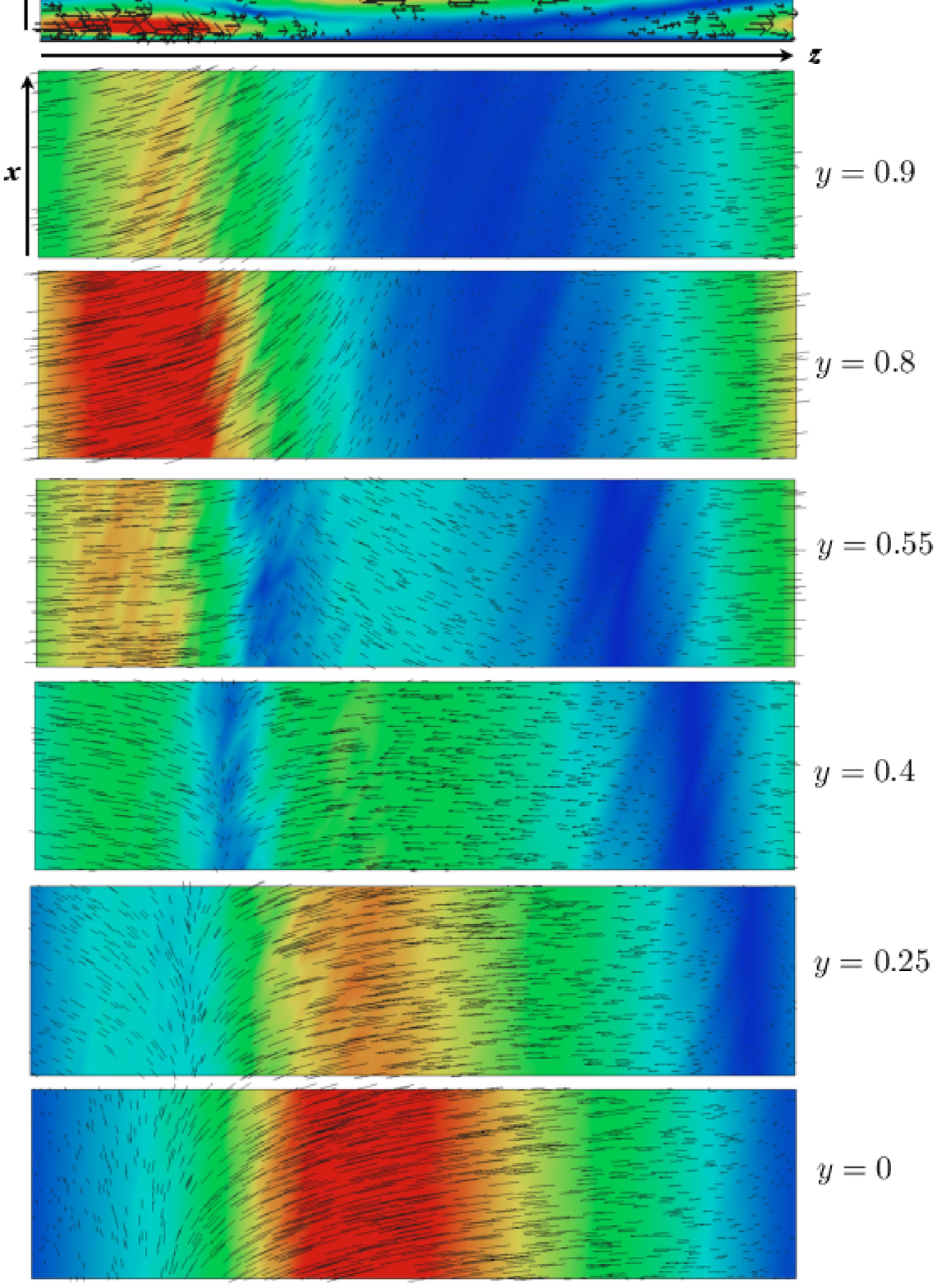}
\caption{Time-averaged deviation from laminar flow at $Re=1100$ on a 
typical $(z,y)$ plane (top) and on $(x,z)$ planes at various values of $y$.  
The $(z,y)$ plot has been stretched by a factor of 3 in the $y$ direction 
for visibility. Flow for $y<0$ resembles that for $y>0$.
Arrows indicate the direction of the velocities while 
colors indicate its magnitude. Scale [0,0.2].}
\label{fig:big_xz_yz}
\end{figure}

\begin{figure*}
\includegraphics[width=17cm]{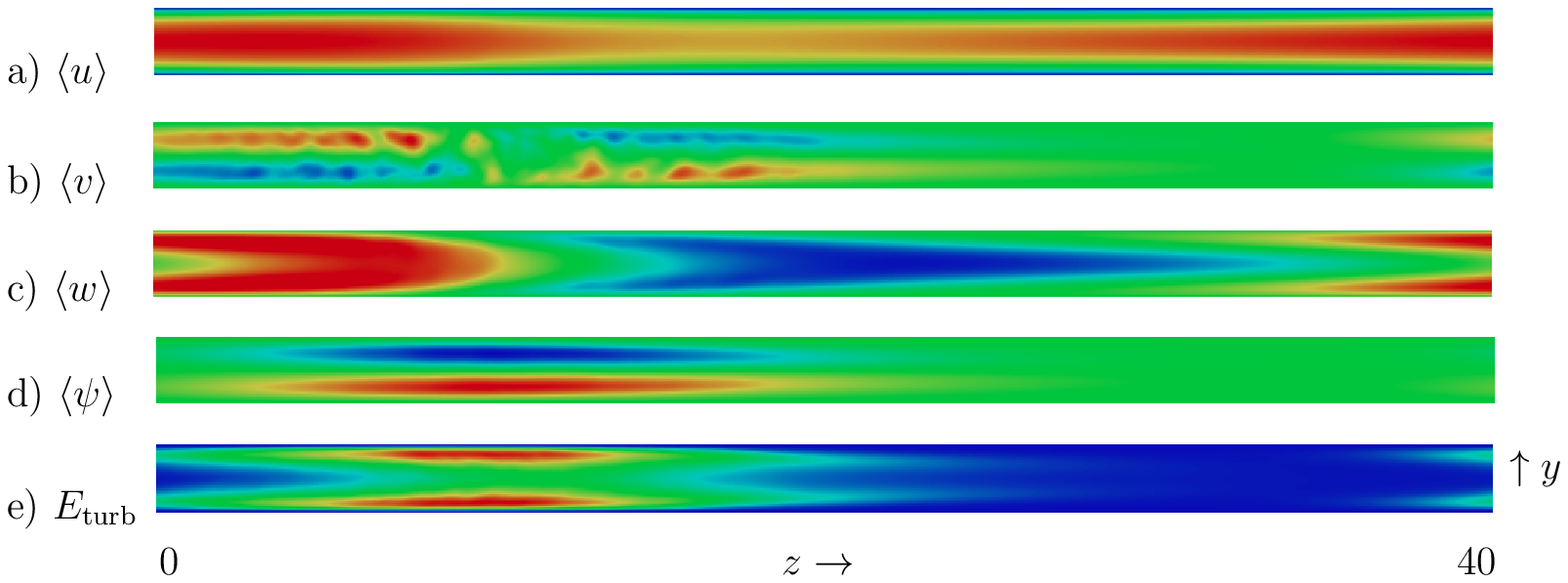}
\caption{Mean flow in the $(y,z)$ plane from simulation at $Re=1100$ 
averaged over $L_x$ and over time of $\Delta T=8000$.
\newline
a) Streamwise velocity $\langle u \rangle$ including laminar profile:
the undulations correspond to profiles which are slug-like 
(tur\-bu\-lent regions)
or parabolic (laminar regions) in Fig.~\ref{fig:time}a.
Scale $[-0.67,0.33]$.
\newline
b) Cross-channel velocity $\langle v\rangle$: small-scale structures 
are still visible despite averaging. 
Scale $[-0.003,0.003]$.
\newline
c) Spanwise velocity $\langle w \rangle$ with characteristic chevrons. 
Scale $[-0.05,0.05]$.
\newline
d) Streamfunction $\langle \psi\rangle$ shows two superposed layers of
cellular flow in the $(y,z)$ plane. The laminar velocity has been subtracted.
\newline
e) Turbulent kinetic energy $E_{\rm turb}$: the red regions show a strong concentration 
near the bounding plates. Scale $[0,0.012]$.}
\label{fig:tsuka_compare}
\end{figure*}

Figure \ref{fig:big_xz_yz} presents views on various planes of the 
deviation of the time-averaged flow from the laminar velocity.
The Reynolds number is 1100, as in Fig.~\ref{fig:3D_view}.
The flow varies a great deal with $y$ and $z$, but depends little on $x$, 
as predicted for the tilted domain. 
In order to gain more insight into this flow, we therefore 
form a 2D field by averaging over $L_x$ as well as $\Delta T=8000$:
\begin{equation}
\langle \bu\rangle (y,z)\equiv \int_{t=0}^{8000} dt \int_{x=0}^{L_x}\bu(x,y,z,t)
\end{equation}
where $\langle\rangle$ has been redefined from (\ref{eq:tavg}). 
Fig.~\ref{fig:tsuka_compare} presents various aspects of $\langle \bu
\rangle$. This figure agrees extremely well with Fig.~3 of 
Tsukahara et al.~\cite{Tsukahara_THMT_06}, which shows similar 
quantities for a patterned flow at $Re=1327$ averaged over time and $L_x$.
Tsukahara et al.~\cite{Tsukahara_THMT_06} do not show the 
spanwise velocity, and include a number of other quantities 
such as the shear stress and Reynolds shear stress not shown here.
The field is reflection-symmetric in $y$. (Reflection in $y$ 
changes the sign of the cross-channel velocity, as it does for the 
streamwise vorticity shown in Fig.~\ref{fig:3D_view}b.)

Figure \ref{fig:tsuka_compare}a shows the streamwise velocity.
Unlike the other parts of Fig.~\ref{fig:tsuka_compare}, 
this subfigure includes the laminar flow. The waviness corresponds 
to the alternation of parabolic and plug profiles which occur 
in laminar and turbulent regions, respectively.
The cross-channel velocity (Fig.~\ref{fig:tsuka_compare}b) 
still shows small-scale features despite the averaging over $L_x$ and
$\Delta T=8000$.
The spanwise velocity (Fig.~\ref{fig:tsuka_compare}c) 
shows distinctive chevron features. 
Figure \ref{fig:tsuka_compare}d depicts the streamfunction 
associated with the deviation of the mean velocity from the laminar flow 
in the $(y,z)$ plane. Two wide counter-rotating cells are stacked in the gap.
This is also the form of the mean flow observed experimentally 
in the presence of a turbulent spot by Lemoult et al.~\cite{Lemoult_EPJE}.
The direction of rotation of these cells is such as to 
slow the streamwise flow in the middle of the channel 
and accelerate it near the walls, in effect transforming 
a parabolic profile to the slug profile. 

\begin{figure*}
\includegraphics[width=17cm]{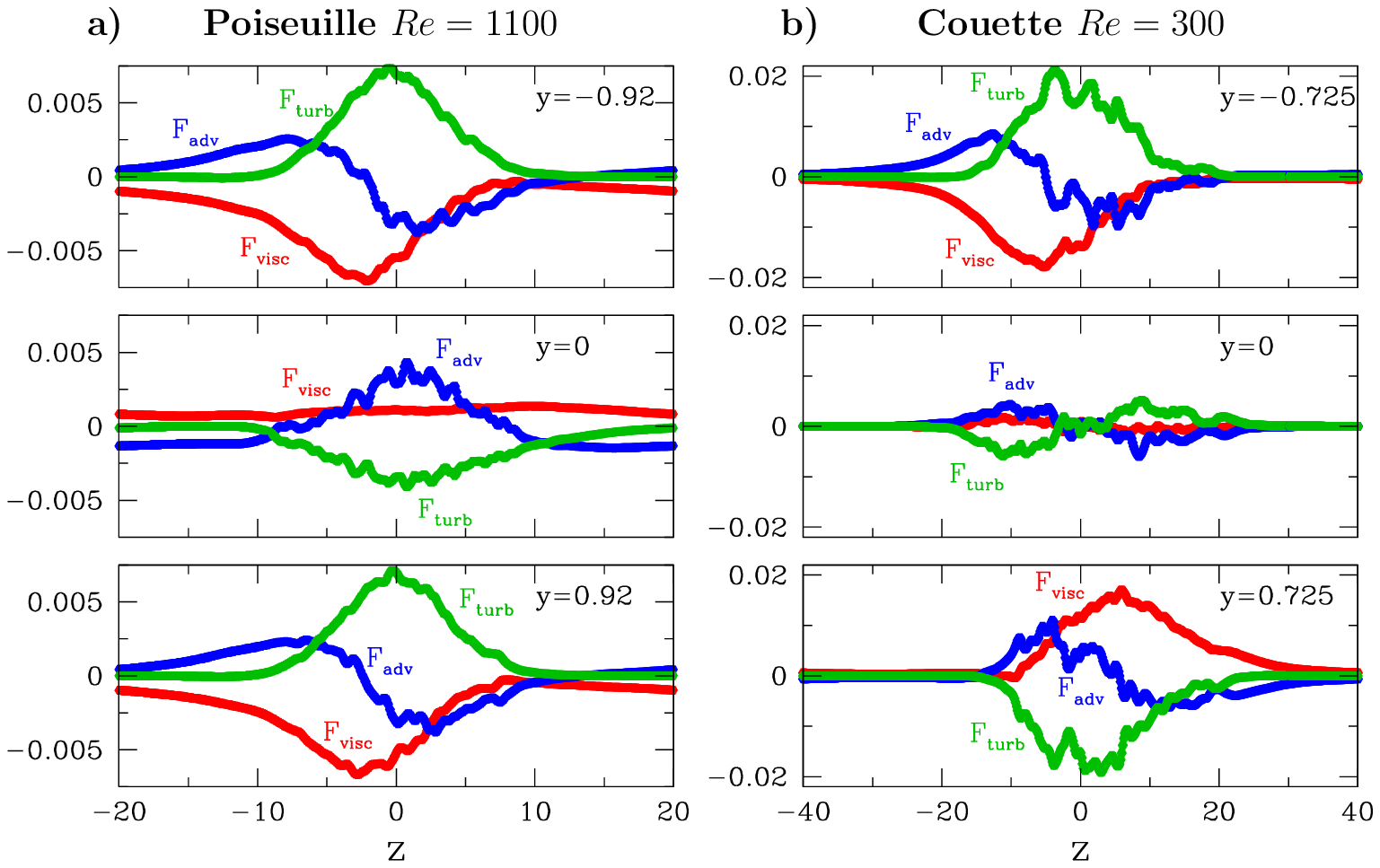}
\caption{Streamwise Reynolds stress (green, \eqref{eq:Fturb}), 
advective (blue, \eqref{eq:Fadv}) and viscous (red, \eqref{eq:Fvisc}) 
forces as a function of $z$ at three $y$ locations for the mean
  flow associated with turbulent-laminar (a) Poiseuille flow at $Re=1100$ 
  and (b) Couette flow at $Re=300$.  Near the lower wall ($y<0$), 
  for both flows, the Reynolds stress force accelerates the fluid through
  the turbulent band while the viscous force decelerates it; advection by the
  basic laminar flow changes sign in the middle of the band. In the center
($y=0$), for Poiseuille flow, advection accelerates the fluid while the 
Reynolds stress force decelerates it; for Couette flow, this situation 
is reversed over half the turbulent region.
Near the upper wall ($y>0$), the force balance for Poiseuille flow is 
identical to that near the lower wall; for Couette flow 
the balance near the upper and lower walls are related by centro-symmetry.}
\label{fig:balance}
\end{figure*}

Figure \ref{fig:tsuka_compare}e presents the turbulent kinetic energy
defined by 
\begin{equation}
E_{\rm turb}\equiv\frac{1}{2}\langle\bufluc\cdot\bufluc \rangle,\qquad
\bufluc \equiv \bu-\langle\bu\rangle
\end{equation}
which is concentrated very near the boundaries, 
where the shear of the laminar profile is greatest. 
The counter-rotating cells (d) are centered at the same value of $z$ 
as the turbulent kinetic energy (e), but the maximum deviation in 
the streamwise velocity (a) is located to the right of this location. 
A shift 
between these quantities was previously noted 
by Tsukahara et al.~\cite{Tsukahara_THMT_06} 
as well as in the case of plane 
Couette flow~\cite{Prigent_PRL,Prigent_PhysD,Barkley_JFM_07,Tuckerman_PF_11}.

We can compare the appearance of these mean quantities 
with the analogous ones in Fig.~5 of Barkley and 
Tuckerman \cite{Barkley_JFM_07} for plane Couette flow. 
As is the case for plane Couette flow, 
variation in $z$ is much slower than variation in $y$,
For plane Couette flow, the turbulent kinetic energy 
occupies most of the interior of the gap and the $(y,z)$ flow 
consists of a single large rotating cell. 
An obvious difference between the two flows is symmetry: 
plane Poiseuille flow is reflection-symmetric in $y$, 
while plane Couette flow is centro-symmetric in $(y,z)$. 
That is, variables $u$ and $w$ obey 
\begin{subequations}
\begin{eqnarray}
F(-y,z)=&F(y,z) \qquad&\mbox{Poiseuille}
\label{eq:refl}\\
F(-y,z)=&-F(y,-z) \qquad&\mbox{Couette}
\label{eq:centro}
\end{eqnarray}
\end{subequations}
For other quantities, e.g.~$v$ or $\psi$, 
the change in sign is opposite to that in 
\eqref{eq:refl} and \eqref{eq:centro}.
One of the striking properties of turbulent-laminar banded 
patterns is that their mean flow inherits the symmetries 
of the laminar flow. 

Figure \ref{fig:balance}a shows the main forces acting in the 
streamwise direction on the mean flow: 
\begin{subequations}
\begin{eqnarray}
F_{\rm turb} &\equiv &-\langle (\bufluc\cdot\grad )\bufluc\rangle \label{eq:Fturb}\\
F_{\rm adv} &\equiv &- \left(\bu_{\rm lam}\cdot\grad\right) \langle u-u_{\rm lam}\rangle \label{eq:Fadv}\\
F_{\rm visc} &\equiv& \frac{1}{Re} \lap\langle u -u_{\rm lam}\rangle
\label{eq:Fvisc}
\end{eqnarray}
\end{subequations}
We omit the larger forces governing laminar Poiseuille flow:
\begin{equation}
\grad p_{\rm lam} = \frac{1}{Re}\lap \bu_{\rm lam} = -\frac{2}{Re} \be_{\rm strm}
\end{equation}
as well as the smaller pressure gradient associated with 
$\langle\bu-\bu_{\rm lam}\rangle$ 
and the nonlinear interaction of $\langle\bu-\bu_{\rm lam}\rangle$ with itself.
The three forces are plotted as a function of $z$ for values of $y$ 
near the two walls and at the center of the channel.
To interpret Fig.~\ref{fig:balance}, it is helpful to 
recall that while $z$ is not the streamwise direction, 
it has a component in this direction; see Fig.~\ref{fig:domain}c. 
Thus a streamwise force which is 
positive (negative) accelerates (decelerates) the fluid towards 
the right (left) in the $z$ direction. 

Figure \ref{fig:balance}b shows the balance of 
forces for an analogous state in plane Couette flow.
We have chosen $Re=300$ because at this Reynolds number, 
the turbulence in plane Couette flow is localized \cite{Barkley_PRL_05}, 
as is the case for plane Poiseuille flow at $Re=1100$.
See section \ref{sec:Discussion} for a discussion 
on converting between scales in Couette and Poiseuille flows.
The domain of length $L_z=80$ is chosen to be twice that 
of the domain we used for Poiseuille flow. 
For each flow, the origin in $z$ has been translated 
so that the turbulent region is at the center of the graph. 
The nonzero $y$ positions for which the forces are plotted 
are those for which the forces are maximal. 
The different symmetries of the two flows, 
i.e.~$y$-reflection for Poiseuille and centro-symmetry in $(y,z)$ 
for Couette flow, are clearly visible in Fig.~\ref{fig:balance}.

The upper panels show the forces for $y<0$, where the shear 
of both basic flows is positive. 
The resemblance between the force balance for 
the two flows is remarkable. For both flows, 
$F_{\rm turb}$ is large and positive in the turbulent region, 
accelerating the fluid towards the right, 
and is counterbalanced primarily by $F_{\rm visc}$. 
The advective force $F_{\rm adv}$ is comparable but smaller in magnitude; 
it acts with $F_{\rm turb}$ in the left portion of the turbulent region 
and against $F_{\rm turb}$ in the right portion. 
At the center ($y=0$), the curvature $\partial_y^2 u$ and hence 
$F_{\rm visc}$ is small for Poiseuille flow and negligible for Couette flow, 
For Poiseuille flow, $F_{\rm adv}$ accelerates the fluid in the 
turbulent region, while $F_{\rm turb}$ decelerates it.
For Couette flow, this holds over half the turbulent region, 
while the reverse is true over the other half, 
as required by \eqref{eq:centro}.
For $y>0$, the flow and forces for Poiseuille flow 
are the same as for $y<0$, while for Couette flow
the flow and forces are reversed from those at $y<0$.

\section{Discussion}
\label{sec:Discussion}

We conclude with some further comparisons between 
turbulent-laminar bands in Poiseuille and Couette flow. 
It has been proposed by Waleffe~\cite{Waleffe_03} that plane Poiseuille flow
can be viewed as two superposed plane Couette flows. This is 
consistent with the two cells and two turbulent regions seen in 
Fig.~\ref{fig:tsuka_compare}d,e. 
The similarity in the balance of forces shown in figure 
\ref{fig:balance} also strongly supports the idea that 
turbulent-laminar patterns are maintained by the same physical 
mechanisms in Poiseuille and Couette flow.

With this in mind, 
we compare the wavelengths and Reynolds numbers of 
turbulent-laminar banded patterns in Poiseuille and Couette flow.
In our domain, the wavelength of the patterns in plane Poiseuille flow 
is 20 at higher $Re$ and becomes 40 for lower $Re$. 
Turbulent-laminar patterns in plane Couette flow have higher wavelengths 
\cite{Prigent_PRL,Prigent_PhysD,Barkley_PRL_05,Barkley_JFM_07,Tuckerman_PF_11}: 
40 for higher $Re$ and 60 for lower $Re$.
The idea of considering Poiseuille flow as two superposed Couette flows 
suggests that Poiseuille flow should be scaled by the quarter-gap 
rather than the half-gap. 
This would make the pattern wavelength of 40 (quarter-gaps) of plane
Poiseuille flow at higher $Re$ consistent with the wavelength of 40 
(half-gaps) observed for plane and Taylor-Couette flow. 
It was already observed \cite{Prigent_PhysD,Barkley_JFM_07} that a unified 
Reynolds number $Re_s$ based on the square of the 
$y$-averaged shear of the laminar flow 
and the quarter gap for plane Poiseuille flow could be defined to yield 
$Re_s= Re_c/4 \approx Re/4.6$. For plane Couette flow, using 
the constant shear and the half-gap, $Re_s$ is the usual Reynolds number.
The range of existence $800 \leq Re \leq 1900$ for turbulent-laminar 
patterns in plane Poiseuille flow becomes $174 \leq Re_s \leq 413$, 
which is of the same order as the range of existence $[300,420]$
for turbulent-laminar patterns in plane Couette flow.
We note that the Reynolds-number range over which patterned turbulence is 
obtained by Tsukahara et al. is $Re_b\in[1125,2250]$, i.e. $Re_s\in[245,490]$ 
in numerical simulations~\cite{Tsukahara_ASCHT_07} 
and $Re_b\in[1275,1500]$, i.e. $Re_s\in[277,326]$ in 
experiments~\cite{Hashimoto_THMT_09}. 
It would be unlikely to obtain more precise agreement
since the analogy between Poiseuille 
and Couette flow is inexact in a number of ways. 
For example, the turbulent near-wall regions of plane Poiseuille flow 
occupy considerably less than a quarter-gap. 
Second, in seeking a single measure of the shear in 
plane Poiseuille flow, it is not clear that a simple average 
is the best candidate. 

An important consequence of the differences in symmetry 
is that the moderate-time averages of the turbulent-laminar patterns in 
Couette flow are stationary while those of Poiseuille flow 
have a well-defined velocity, which we have shown in Fig.~\ref{fig:velocities}b).

It is almost surely possible to produce patterns at angles quite different 
from $24^\circ$. In large-scale experiments in plane Couette 
flow~\cite{Prigent_PRL,Prigent_PhysD}
patterns were observed whose angles ranged between $24^\circ$ and $37^\circ$;
simulations in narrow tilted domains with imposed angles 
ranging from $15^\circ$ and $66^\circ$ all produced 
patterns~\cite{Barkley_PRL_05,Barkley_JFM_07,Tuckerman_PF_11}.
The simulated patterns whose angles are far outside the range 
$[24^\circ,37^\circ]$ would presumably be unstable when placed in a 
less constrained geometry. 
Although the narrow tilted geometry -- the analogue of the 
minimal flow unit~\cite{Jimenez_91,Hamilton}
for maintaining shear-flow turbulence -- can be used 
to study some of the characteristics of turbulent-laminar patterns, 
studies in a less constrained geometry are necessary for understanding 
their genesis and fate. The spreading of turbulent spots 
and fronts have been widely studied for plane Poiseuille flow, 
e.g.~experimentally by Lemoult et al.~\cite{Lemoult_EPJE,Lemoult_JFM_13}
and numerically by Aida et al.~\cite{Aida_TSFP_11} and 
as well as numerically by Duguet et al.~\cite{Duguet_PRE_11} 
for plane Couette flow. 

Future work will focus on the mechanism maintaining turbulent-laminar 
patterns and on the branching events that accompany the change in 
wavelength and in speed. 

\begin{acknowledgments}
This work was performed using high performance computing resources provided by
the Grand Equipement National de Calcul Intensif-Institut du D\'eveloppement
et des Ressources en Informatique Scientifique project 1119.
The authors acknowledge financial support by the \emph{Nieders\"achsisches Ministerium f\"ur Wissenschaft und Kultur}.
Dwight Barkley is acknowledged for valuable discussions on channel and pipe
flow. 
\end{acknowledgments}


\end{document}